\documentclass[amsmath,amssym,amsfonts,pre,twocolumn,showpacs]{revtex4}
\usepackage[dvips]{epsfig}

\newcommand{\av}{{\mathbf a}}

\newcommand{\tv}{\hat{{\bf t}}}
\newcommand{\bv}{{\bf b}}
\newcommand{\xv}{{\mathbf x}}
\newcommand{\uv}{{\mathbf u}}
\newcommand{\rv}{{\mathbf r}}

\begin{document}

\title{Defects in Crystalline Packings of Twisted Filament Bundles:  II. Dislocations and Grain Boundaries}

\author{Amir Azadi}
\affiliation{Department of Physics, University of Massachusetts, Amherst, MA 01003, USA}

\author{Gregory M. Grason}
\affiliation{Department of Polymer Science and Engineering, University of Massachusetts, Amherst, MA 01003, USA}

\begin{abstract}
Twisted and rope-like assemblies of filamentous molecules are common and vital structural elements in cells and tissue of living organisms. We study the intrinsic frustration occurring in these materials between the two-dimensional organization of filaments in cross section and out-of-plane interfilament twist in bundles.   Using non-linear continuum elasticity theory of columnar materials, we study the favorable coupling of twist-induced stresses to the presence of edge dislocations in the lattice packing of bundles, which leads to a restructuring of the ground-state order of these materials at intermediate twist.  The stability of dislocations increases as both the degree of twist and lateral bundle size grow. We show that in ground states of large bundles, multiple dislocations pile up into linear arrays, radial grain boundaries, whose number and length grows with bundle twist, giving rise to a rich class of ``polycrystalline" packings.
\end{abstract}
\pacs{}
\date{\today}

\maketitle

\section{Introduction}   

Topological defects populate the ground states of many frustrated systems in condensed matter physics~\cite{nelsonbook, sadoc_frustration}.    A key example occurs when crystalline order forms on two-dimensional surfaces of non-zero curvature, where the incompatibility between globally straight and parallel directions generates geometrically-induced stresses that favor defects in the crystalline order.  On spheres, where topology requires a minimal number of twelve 5-fold disclinations, the problem of determining the ideal structure of in-plane order is known alternately as the {\it Thomson} or {\it Thames problem}~\cite{altschuler, aste_weaire}.  This problem has important connections to the structure of viral capsids, which are closed-shell assemblies of proteins~\cite{caspar_klug, zandi}, and more recently has been the subject of experimental interest in the context of particle-stabilized emulsion droplets~\cite{dinsmore, bausch}.  A clear physical picture of the coupling between surface curvature and the presence of topological defects has emerged based on the continuum elasticity theory of crystalline membranes~\cite{nelson_peliti, nelson_seung, bowick_travesset_nelson, bowick_giomi}.  In this theory, disclinations carrying a discrete topological charge act as point sources for in-plane stress that can be screened by a more homogeneous distribution of ``topological charge" generated by the appropriate Gaussian curvature of the membrane.  From this viewpoint, the net topological charge of disclinations in the lowest energy states of curved membranes is rationally expected to increase with the integrated Gaussian curvature of a membrane, a prediction which has recently been tested experimentally for 2D crystals on surfaces of both positive and negative curvature~\cite{irvine_nature}.  

Recently, we have shown that frustration of crystalline order on spherically-curved surface is fundamentally connected to frustrated order in a distinct class of two-dimensionally ordered materials, namely, twisted filament assemblies~\cite{grason_prl_10, grason_pre_11}.  Twisted assemblies of fibrous proteins are common and important structural elements in many biological materials, such collagen~\cite{fratzl, wess} and fibrin~\cite{weisel}.  In these assemblies, helical twist of the assembly derives from the nature of interactions between chiral biofilaments, while the dense in-plane packing results from strong cohesive interactions between filaments~\cite{weisel_pnas, turner, grason_prl_07, grason_pre_09, hagan_10, heussinger}.  Unlike the case of crystalline membranes where frustration arises from out-of-plane deflections, the frustration of cross-sectional order in the twisted-filament bundle derives from a unique geometrical coupling of in-plane strains and filament tilts~\cite{grason_prl_10}.  Despite the distinct geometrical origin, twist generates stresses in the cross-section of filament bundles that are formally equivalent to those induced by a {\it positive Gaussian curvature} in a membrane, corresponding to a spherical geometry of effective radius, $R_{eff} = \Omega^{-1} /\sqrt{3}$, where $2 \pi/\Omega$ is the pitch of helical bundle twist.  Strictly speaking, due to the free surface at the boundary of the bundle, the twisted-filament packing maps more closely onto the problem of crystalline order of a partial, spherical cap, a problem that has be studied theoretically for both the cases with~\cite{giomi_bowick} and without~\cite{gompper} topological defects.  Based on this connection, in previous work~\cite{grason_prl_10, grason_pre_11} it was argued that the ground-state order of filament bundles becomes unstable to one or more {\it 5-fold disclinations} in the cross sectional order, when the twist is greater than a critical value, $|\Omega R|_c = \sqrt{2/9} \simeq 0.47$, where $R$ is the bundle radius.  Therefore, a range of multi-disclination ground states were predicted for sufficiently large and twisted bundles.

In this paper, we study the continuum elasticity theory of twisted bundle cross sections to explore a fundamentally distinct class of topological defect configurations in the ground states:  ``neutral" 5-7 disclination pairs, or edge dislocations in the cross-sectional packing.  Though the positive topological charge of a bare 5-fold disclination best neutralizes the negative effective charge generated by the twist, we find a broad range of conditions for which configurations of  5-7 dipoles, or dislocations, in the bundle cross section provide a lower-energy means of screening the geometrically-induced costs of twist.  Interestingly, we find that appropriately polarized dislocations are universally attracted to a radial position at $R/\sqrt{3}$ away from the bundle center, a point associated with vanishing azimuthal stresses in twisted bundles.    In untwisted bundles, as in a flat 2D crystal, edge dislocations incur an elastic cost per unit volume of roughly $K_0 |\bv|^2 \ln (R/|\bv|)$, where $\bv$ is the Burger's vector, $K_0$ is the 2D Young's modulus~\cite{halperin_nelson}.  In twisted bundles, we show that an elastic coupling between geometrically-induced stress and dislocations leads to an additional elastic energy gain per unit volume for optimaly-placed dislocations proportional to $- K_0  \Omega^2 R |\bv|$.  Thus, we show that twisted bundles become unstable to edge dislocations for reduced bundle twists above a threshold value, $|\Omega R| > (\Omega R)_* \sim a/R \ln (R/a)$, where $a \approx |\bv|$ is the lattice spacing of the bundle.  Unlike the case of disclination stability~\cite{grason_pre_11}, find that the stability of dislocations is governed by both the twist of the bundle, $(\Omega R)$, as well as the size of the bundle relative to the microscopic size of filaments, $R/a$.  Importantly, this analysis shows that the critical twist for dislocation stability,  $(\Omega R)_*$, decreases, albeit slowly, to zero as the bundle grows macroscopic in size, as $R/a \to \infty$.  A key consequence of this analysis is that large bundles are unstable to {\it dislocations} over a range of intermediate twist before {\it 5-fold disclinations} are favored for $|\Omega R| >(\Omega R)_c$. 

\begin{figure}
\center \epsfig{file=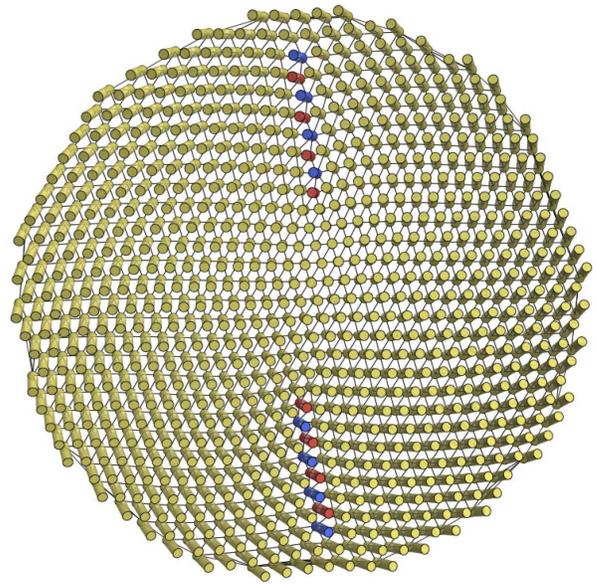, width=3.25in}
\caption{Cross-sectional view of the microscopic model of a helical filament bundle with two small-angle grain boundary arms, each with four dislocations. Dislocations are polarized such that 5-fold dislinations (red) are closer to the bundle center than 7-fold (blue). }
\label{fig: lattice}
\end{figure}

Within this intermediate range of twist $(\Omega R)_* < |\Omega R| <  (\Omega R) _c$, we predict a range of complex multi-dislocation ground states, which are quite distinct in structure from well-separated defects in multi-disclination packings of twisted bundles studied in ref. \cite{grason_prl_10}.   In these ground states multiple 5-7 dipoles form extended, linear chains, or {\it grain boundaries}, aligned along the radial directions.    Grain boundaries in twisted bundles run from the free surface of bundles and terminate before reaching the central core.  An example of a two grain boundary packing of twisted bundles is shown in Figure \ref{fig: lattice}.  The multi-dislocation ground states of twisted bundles are quite similar to the ``grain boundary scars" predicted~\cite{bowick_travesset_nelson, bowick_caccuito} and observed~\cite{bausch, irvine_nature} on spherical crystals in which neutral 5-7 pairs decorate the twelve topologically-required 5-fold disclinations on the sphere.   Despite the similar affinity of twist-induced stresses in bundles for positively-charged disclinations, at intermediate levels of twist, the energetically favored defects arrays are ``neutral", possessing no excess of 5-fold disclination charge.  Based on numerical and scaling analysis of grain boundary screening of twist induced stress we deduce the dependence of the total number of dislocations, $N_d$, and number of grain boundaries, $M$,  on twist and bundle size.  Far above the critical twist we find that $N_d \sim (R/a )(\Omega R)^2$ and $M \sim N_d$.  

The organization of this article is as follows.  In Sec. \ref{sec: continuum} we review the continuum theory of two-dimensionally ordered filament arrays.  In Sec. \ref{sec: effect} we derive the effective energy of dislocations in the cross sections of twisted bundles, and determine the stability of neutral (dislocations) and charged (disclinations) topological defects in the ground states of twisted bundles.  In Sec. \ref{sec: multi} we analyze the structure and thermodynamics of multi-dislocation ground states and appearance of multiple grain boundaries in the cross sections. We also exploit a scaling argument based on the geometry of small-angle grain boundaries to establish the quantitative connection between the length and number of defect arrays in the ``polycrystalline" ground states of twisted bundles.

\section{Continuum theory of filament bundles}

\label{sec: continuum}

The derivation of the equations of mechanical equilibrium in the non-linear continuum elasticity theory of twisted filament bundles have been presented in detail previously~\cite{grason_pre_11}.  In this section we briefly review the key elements of this analysis, which allow us to construct the effective theory of dislocations in the next section.

We consider a cylindrical bundle of radius, $R$,  of filaments of unlimited length.  The stress-free reference state is a hexagonal packing of the straight filaments in the cross section.  We describe the elastic cost of deformations of the cross section order by the following energy,
\begin{eqnarray}
\label{eq: energy}
E=\frac{1}{2}\int dV( {\lambda u^{2 }_{kk}+2\mu u_{ij}u_{ij}}) .
\end{eqnarray}
Here, $\lambda$ and $\mu$ are the Lam\'e elastic coefficients characterizing the elastic properties of the material and correspond to compressive and shear distortion of the array respectively, and $u_{ij}$ is the 2D strain of cross sectional order, defined below.  In this theory, the elastic energy will penalize distortions of the array that change distances between filaments in a plane {\it locally perpendicular to the filament tangent direction}, $\tv$.   Hence, eq. (\ref{eq: energy}) describes the elastic response of hexagonal-columnar material~\cite{degennes_prost}.    While the filaments are uniformly aligned along the $\hat{z}$ direction in the initial configuration, displacements of the array in general lead to filaments that are tilted into the $xy$ plane of initially perfect hexagonal order~\cite{selinger}.  This relationship is captured by introducing a two-component displacement field, ${\bf u} (\xv) = \rv_\perp(\xv) - \xv_\perp$, describing the local deviation in the $xy$ plane of a filament initially at $\xv$ and displaced to a position $\rv(\xv)$.  For small strains, the in-plane displacement is related to the filament tilt geometrically by, $\tv(\xv) = \hat{z} + \partial_z \uv$.  

To satisfy the symmetry considerations described above, the non-linear strain tensor has the form
\begin{eqnarray}
u_{ij}=\frac{1}{2}(\partial_{i}u_{j}+\partial_{j}u_{i}+\partial_{i} \uv \cdot \partial_{j} \uv -\partial_{z}u_{i}\partial_{z}u_{i}),
\end{eqnarray}
which, like the displacement, only has components in the $xy$ plane.  The first two terms on the right hand side are the standard symmetric derivatives in the elastic strain tensor.  Additionally, there are two non-linear contributions to the strain tensor.  The third term ensures the rotational invariance of the 2D solid around the $\hat{z}$ axis.  The final term is unique to the theory of columnar materials and preserves the invariance of the elastic energy about an axis in $xy$ plane \cite{selinger}.   Since $\tv_\perp \simeq \partial_z \uv$, intuitively this contribution to $u_{ij}$ shows that for a fixed separation in the $xy$ plane, when neighboring filaments are tilted with respect to each other, the distance of closest approach between them is reduced.  The presence of this non-linear coupling between filament tilts and in-plane strain necessarily introduces stress in twisted filament bundles~\cite{grason_prl_07}.  

In this study, we consider helically-twisted filament bundles in which the cross-sectional positions of filaments are reorganized due to the presence of geometrically-induced stresses.  Formally, we compose the displacement field of the helically twisted assembly from two deformations:  an initial ($z$-invariant) in-plane displacement, $\uv(\xv)$, followed by a uniform helical twist, at a rate $\Omega$, around the $\hat{z}$ axis.  We denote the composite ($z$-dependent) displacement as $\uv_\Omega$, which has the form
\begin{multline}
\mathbf{u}_{\Omega}(\xv)=\cos(\Omega z)\big[(x+u_x)\mathbf{\hat{x}}+(y+u_y)\mathbf{\hat{y}}\big] \\ -\sin(\Omega z)\big[(y+u_y)\mathbf{\hat{x}}-(x+u_x)\mathbf{\hat{y}}] - \xv_\perp .
\end{multline}
In this configuration is it is straightforward to show that in-plane components of filament orientation have the following texture,
\begin{eqnarray}
\mathbf{\hat{t}_{\perp}}\simeq \partial_{z} \mathbf{u}=\Omega \rho \hat{\phi},
\end{eqnarray}
where $\rho$ is the radial distance of the filament from the bundle center in the deformed state, and $\hat{\phi}$ is the azimuthal direction, also defined with respect to the deformed, or ``current" position of the filament.  

The helical symmetry of these configurations allow us to describe the state of strain for all $z$, based on the in-plane displacement field, $\uv(\xv)$ at $z=0$.  We assume the rate of twist and filament orientation, described by displacement $\mathbf{u}_{\Omega}$, eq. (3), to be fixed and allow for mechanical equilibrium by relaxing the in-plane displacements $\mathbf{u}$. This assumption has the advantage that it reduces the problem energy minimization to one of 2D elasticity theory.   Minimization of the elastic energy, eq. (\ref{eq: energy}), with respect to variations in $\uv(\xv)$ leads to the Euler-Lagrange equations that describe the static mechanical equilibrium of the system,
\begin{equation}
\label{eq: EL}
\frac{\delta(E/L)}{\delta u_{i}}\simeq -\partial_{j}\sigma_{ij}= 0
\end{equation}
where $L$ is the length of the bundle and the stress tensor has the standard form an isotropic, 2D elastic medium, $\sigma_{ij}=\lambda u_{kk}\delta_{ij}+2\mu u_{ij}$.  Here, as in ref. \cite{grason_pre_11} we have neglected a term $t_j \sigma_{jk} \partial t_k /\partial r_j$ from eq. (\ref{eq: EL}) because it contributes to the stress balance of twisted bundles at higher order in reduced twist, $\Omega R$, which is assumed to be smaller than unity.     As the surface of the bundle is free to move, we solve for states of mechanical equilibrium subject to a vanish normal stress at the boundary of bundle
\begin{eqnarray}
\hat{r}_{i}\sigma_{ij}(\rho=R)=0.
\end{eqnarray}
We proceed to solve for the divergence-free stress, in terms of the Airy stress function $\chi$ \cite{landau_lifshitz}, related to the stress tensor by,
\begin{eqnarray}
\sigma_{ij}=\epsilon_{ik}\epsilon_{j\ell }\partial_{k}\partial_{\ell}\chi .
\end{eqnarray}
While this definition of $\sigma_{ij}$ satisfies eq. (\ref{eq: EL}) by construction, it is necessary to enforce extra conditions on $\chi$ that ensure that the stress corresponds to the physical configuration of $\uv$.  As in ref. \cite{nelson_seung}, this compatibility relation may be derived be equating the anti-symmetric derivatives of strain, $\epsilon_{ik} \epsilon{j \ell} \partial_k \partial_\ell u_{ij}$, from which we derive
\begin{eqnarray}
\label{eq: compat}
K_0^{-1} \nabla_\perp^4 \chi = s(\xv) + \nabla_\perp \times \bv (\xv) -K_T .
\end{eqnarray}
where $K_{0}=4\mu (\lambda+\mu)/(\lambda+2\mu)$ is the 2D Youngs modulus.  The right-hand side of eq. (\ref{eq: compat}) may be viewed as sources for Airy stress.  The first and second of these denote the sources of stress generated by topological defects in a bundle cross section, disclinations and dislocations respectively, for which the solution for $\uv(\xv)$ is not single-valued.   The final term, denoted as the {\it intrinsic twist}, derives from the non-linear contribution to $u_{ij}$ from filament tilt,
\begin{equation}
K_T=\frac{1}{2}\epsilon_{ik}\epsilon_{j \ell}\partial_{k}\partial_{\ell}t_{i}t_{j}=3\Omega^2 .
\end{equation}
A similar contribution derives from the non-linear coupling of elastic strain and membrane tilt in the continuum theory of elastic membranes, where it is known that minus the Gaussian curvature of the membrane acts a source for Airy stress~\cite{nelson_seung}.  Hence, we see that twist in filament bundles generates in-plane stresses that are formally equivalent to those generated by spherical membranes of curvature $3 \Omega^2$.  

In general, two types of defects contribute to the right-hand side of equation (\ref{eq: compat}) as the sources of the stress: disclinations and dislocations~\cite{nelson_seung}. Disclinations are disruptions of the orientational symmetry of the lattice and are associated with singular configurations of $\theta_6 (\xv)=\frac{1}{2}\epsilon_{ij}\partial_{i}u_{j}$, the bond angle of the lattice.  Around a single disclination, $\theta$ increases or decreases by an integer multiple of $2\pi/6$,
\begin{eqnarray}
\oint d\ell \cdot \nabla_{\perp}\theta_6=s,
\end{eqnarray}
where $s=(2\pi/6)n$ is the topological charge of the disclination.  Dislocations are associated with singular configurations of $\uv(\xv)$ and defined in terms of a closed loop integral around which $\mathbf{u}$ changes by an integer multiple of the lattice spacing along one of the six-fold directions,
\begin{eqnarray}
\oint d\ell \cdot \nabla_{\perp}u_{i}=b_{i},
\end{eqnarray}
 where $\mathbf{b}$ is the Burgers vector.  Multiple point defects in the cross section correspond to the defect densities, $s(\xv) = \sum_\alpha s_\alpha \delta^{(2)} (\xv-\xv_\alpha)$ and $\bv(\xv) = \sum_\alpha \bv_\alpha \delta^{(2)} (\xv-\xv_\alpha)$. 

In the presence of twist- and defect-induced stresses, eq. (\ref{eq: compat}) may be solved for $\chi$ and subsequently the elastic energy may be computed from,
\begin{equation}
E = \frac{1}{2 K_0} \int d V (\nabla^2_\perp \chi)^2 .
\end{equation}
In ref. \cite{grason_pre_11} these equations were solved in the presence of an arbitrary array of disclinations in the cross section of filament bundles by multi-pole expansion, yielding an effective energy written purely in terms of charge and position of disclinations and bundle twist
\begin{multline}
\label{eq: disc}
 \frac{E}{V K_0} = \frac{ 3 (\Omega R)^4}{128} + \sum_\alpha \frac{s_\alpha}{32 \pi} \Big[ \frac{s_\alpha}{ \pi} - \frac{ 3 (\Omega R)^2}{2} \Big] \Big(1- \frac{\rho_\alpha^2}{R^2} \Big) \\
 + \frac{1}{2} \sum_{\alpha \neq \beta} s_\alpha V_{int}(\xv_\alpha, \xv_\beta) s_\beta ,
 \end{multline} 
where $V$ is the bundle volume and 
\begin{multline}
\label{eq: discint}
V_{int} (\xv_\alpha, \xv_\beta) = \frac{1}{16 \pi^2} \Big(1- \frac{\rho_\alpha^2}{R^2} \Big)\Big(1- \frac{\rho_\beta^2}{R^2} \Big) \\
+\frac{ |\Delta \xv_{\alpha \beta}|}{16 \pi^2 R^2} \ln \bigg[ \frac{  |\Delta \xv_{\alpha \beta}|^2}{ (R^2- \rho_\alpha^2) (R^2- \rho_\beta^2)/R^2 +  |\Delta \xv_{\alpha \beta}|^2} \Bigg] ,
\end{multline}
and $\Delta \xv_{\alpha \beta} = \xv_\alpha - \xv_\beta$.  This energy has three contributions: the first term describes the elastic cost of twist; the second term is the defect self energy and twist-defect interaction; and the third term describes the elastic interaction between disclinations. Importantly, both the disclination self-energy terms in (\ref{eq: disc}) and interaction terms in (\ref{eq: discint}) vanish continuously as disclinations approach the bundle surface, $\rho_\alpha \to R$.  As noted in \cite{grason_pre_11}, this property derives from the screening of far-field stresses induced by topological defects by boundary-induced stresses as defects draw near to the free boundary.

\section{Elastic energy of dislocations in twisted bundles}

\label{sec: effect}

\subsection{Dislocation energies and interactions}
 
In this section, we take advantage of the dual description of dislocations, which may be constructed from neutral 5-7 pairs of disclinations~\cite{halperin_nelson} to derive the continuum theory of dislocation energies and interactions in twisted bundles.  The theory of edge dislocations in the cross section of (untwisted) cylindrical crystals was originally studied in detail by Koehler~\cite{koehler}.  In this study, the resulting forms for dislocation self-energy and interaction energies were derived in terms complex area-integrals of stress distribution overlap, which were then analyzed numerically.  In the present study, the exact, closed-form expressions for {\it disclinations energies} derived in ref. \cite{grason_pre_11} allow us to derive the algebraic formula for the full position- and orientation dependence of {\it dislocation} energies in cylindrical crystals.  

\begin{figure}
\center \epsfig{file=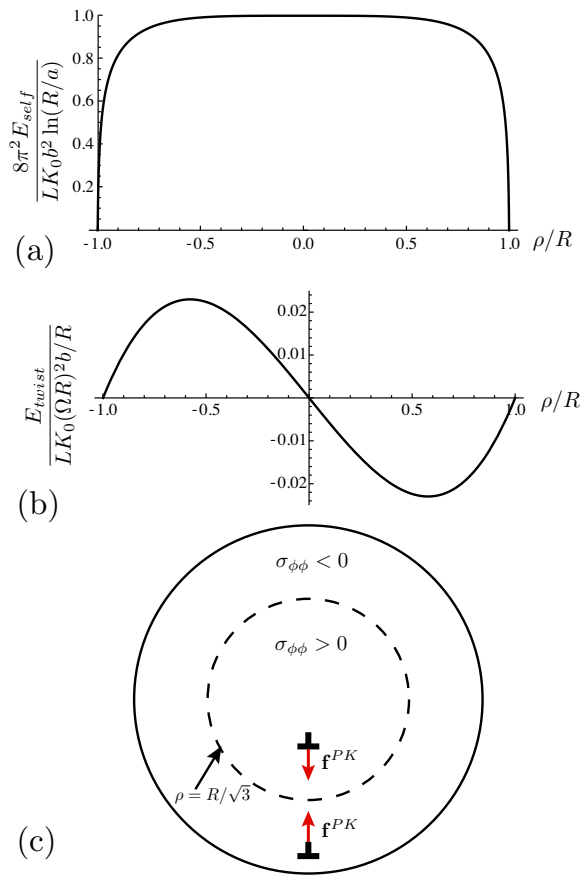, width=3.in}
\caption{The dislocation self-energy is plotted in (a) vs. radial position, $\rho$.  In (b), we show the radial dependence of the elastic coupling between twist and dislocation stresses.   In (c), a schematic showing the Peach-Koehler force on disclocations in the presence of twisted induced stress, where ${\bf \bot}$ indicates the position of an edge dislocation.  The dashed line indicates a contour of vanishing hoop stress, to which dislocations in highly-twisted bundles are driven.  For each figure, the dislocation orientation is $\bv = b \hat{\phi}$.}
\label{fig: energy}
\end{figure}

From eq. (\ref{eq: compat}) we may show that far field stresses generated by a single dislocation of Burgers vector $\bv$ at $\xv$, may be constructed by superposing a 5-fold disclination, $s=+2\pi/6$, at $\xv+\av/2$ and 7-fold disclination, $s=-2\pi/6$, at $\xv - \av/2$, where $ \hat{z} \times \av =  (2\pi /6 )\bv$.  Defining $\chi_{+}(\mathbf{x})$ as the Airy stress generated by a single 5-fold disclination, $s= + 2 \pi/6$, which was calculated exactly in ref. \cite{grason_pre_11}, the Airy stress corresponding to a dislocation at $\xv$, denoted by $\chi_{disl}(\xv)$, is given by 
\begin{equation}
\chi_{disl}(\xv) = |\bv| \lim_{a \to 0} \Big[ \frac{ \chi_{+}(\xv+\av/2)- \chi_{+}(\xv-\av/2) }{a} \Big] .
\end{equation}
To calculate the energy of a single edge dislocation in the bundle cross section, we simply superpose a 5-7 disclination pair separated by a vector $\av$, sum the self- and interaction energies described in eq. (\ref{eq: disc}), and expand the resulting energy to second order in $a/R$.  This results in the following energy for a single dislocation,
\begin{equation}
\label{eq: Edisl}
E_{disl} = E_{self}+ E_{twist}.
\end{equation}
where,
\begin{multline}
\label{eq: Eself}
\frac{E_{self}} {V K_0} =  \frac{b_\phi^2}{8\pi^{2}R^{2}}\left(\frac{\rho}{R}\right)^{2}\\+\frac{|\bv|^2}{8\pi^{2}R^{2}}\bigg[ \ln{\Big(1-\frac{\rho^{2}}{R^{2}}\Big)}+\ln{\left(\frac{R}{a}\right)}\bigg]
\end{multline}
and
\begin{equation}
\label{eq: Etwist}
\frac{E_{twist} }{V K_0} =-\frac{3\Omega^{2}}{16\pi}b_{\phi}\rho\left(1-\frac{\rho^{2}}{R^{2}}\right) .
\end{equation}
$E_{self}$ is the elastic energy of a single dislocation in an untwisted bundle, which depends largely on the radial position, $\rho$, and weakly on orientation of the dislocation.  This energy is maximal for a central dislocation, $\rho =0$, and reduces to the well known logarithmically divergent cost for a single dislocation in a bulk crystal~\cite{nabarro, hirth}.  The radial dependence of $E_{self}$ shown in Fig. \ref{fig: energy}a, becomes singular as the dislocation approaches the bundles surface as the boundary-induced force on a dislocation diverges as $~\sim (R-\rho)^{-1}$~\cite{koehler}.  Hence, in the limit that $R-\rho \ll a$, the 5-7 disclination superposition is non-analytic as $a \to 0$, and hence the small-$a$ expansion of eq. (\ref{eq: Eself}) becomes inaccurate.

$E_{twist}$ describes the elastic coupling between the twist induced stresses and the dislocation stresses shown in Fig. \ref{fig: energy}b.  Notably, this coupling is negative and minimal for $\bv= b \phi$ and $\rho= R/\sqrt{3}$, demonstrating that twist favors dislocations of a certain polarization and located at a specific radial position in twisted bundles.  The favorable orientations of dislocations correspond to disclination dipoles oriented along the radial direction, with the 5-fold end oriented towards the bundle center.  Alternatively, we may view such a dislocation as a partially removed row of filaments extending from the free bundle surface to the dislocation.

To understand the origin of an optimal location of dislocation in twisted bundles, we consider an alternative derivation of $E_{twist}$ based on the Peach-Koehler force~\cite{peach_koehler} generated by twist-induced stress.  The force per unit length on a dislocation line along $\hat{z}$ subject to imposed stress $\sigma_{ij}$ is given by,
\begin{equation}
\label{eq: PK}
f^{PK}_i = \epsilon_{ij} \sigma_{jk} b_k .
\end{equation}
The stresses generated by twist are described by the solution to $K_0^{-1} \nabla_{\perp}^{4}\chi_{twist}=-3 \Omega^2$, which can be readily solved to show the following azimuthal stress distribution,
\begin{equation}
\label{eq: sigtwist}
 \sigma_{\phi \phi}^{twist} (\rho)= \frac{3 K_0 \Omega^2}{16} (R^2- 3 \rho^2) .
 \end{equation}
This stress distribution divides the bundle into two regions:  tensile hoop stresses, $\sigma_{\phi \phi}^{twist} >0$, at the bundle core for $\rho < R/\sqrt{3}$; and due to large azimuthal tilt of filaments at the periphery, compressive hoop stresses, $\sigma_{\phi \phi}^{twist} < 0$, for $\rho > R/\sqrt{3}$.  Since twist induces a radially symmetric stress, $\sigma_{\phi r}^{twist}=0$, the force of a dislocation whose Burgers vector is oriented along $\phi$ (with a 5-7 dipole along $\hat{r}$) is in the $\hat{r}$ direction.  As shown in Fig \ref{fig: energy}c, for such a defect in the compressive zone, for $ \rho > R/\sqrt{3}$, the Peach-Koehler force drives the dislocation inwards, while in the tensile zone, for $ \rho < R/\sqrt{3}$, this force drive the dislocation outwards.  Hence, the force vanishes where $ \sigma_{\phi \phi}^{twist}$ vanishes at $\rho= R/\sqrt{3}$, the stable position.  Combining eqs. (\ref{eq: PK}) and (\ref{eq: sigtwist}), energetic coupling between twist and dislocations, eq. (\ref{eq: Etwist}), may be readily calculated from the mechanical work of driving a defect into the bundle, $E_{twist} = -b_\phi \int_\rho^R d\rho'  \sigma_{\phi \phi}^{twist} (\rho ')$.

We end this section with an analysis of dislocation-dislocation interactions in cylindrical bundles.  As with $E_{self}$ and $E_{twist}$, we derive these from the interactions between two neutral disclination pairs.  To compute inter-dislocation energies, we sum disclination interactions over two disclinations, $s_{1}^{\pm}=\pm 2\pi/6$ at $\mathbf{x}_{1}^{\pm}=\mathbf{x}\pm \av_{1}/2$, and the second pair of disclinations, $s_{2}^{\pm}=\pm 2\pi/6$ at $\mathbf{x}_{2}^{\pm}=\mathbf{x}\pm \av_{2}/2$.  Again, we retain terms to lowest order in $a$ from the expansion of multiple disclination interactions, yielding the interaction energy
\begin{widetext}
\begin{multline}
\label{eq: Eint}
\frac{E_{int}}{K_{0}V}=\frac{1}{4\pi^{2}R^{2}}\bigg[-(\mathbf{b}_{1} \cdot \mathbf{b}_{2})\left(\ln \cos ^{2}\xi+\sin^{2}\xi\right) +\frac{(\mathbf{r}_{1}\times\mathbf{b}_{1}) (\mathbf{r}_{2}\times\mathbf{b}_{2})}{R^{2}}\sin^{4}\xi 
+\frac{(\mathbf{b}_{1}\times\mathbf{\Delta x}_{12}) (\mathbf{b}_{2}\times\mathbf{\Delta x}_{21})}{|\mathbf{\Delta x}_{12}|^{2}}\left(1-\cos^{4}\xi\right) 
\\+\frac{(\mathbf{b}_{1}\times\mathbf{\Delta x}_{12}) (\mathbf{b}_{2}\times\mathbf{r}_{2})(1-\rho_1^2/R^2)+(\mathbf{b}_{2}\times\mathbf{\Delta x}_{21})(\mathbf{b}_{1}\times\mathbf{r}_{1}) (1-\rho_2^2/R^2)}{(R^2-\rho_{1}^{2})(R^2-\rho_{2}^{2})+|\mathbf{\Delta x}_{12}|^{2}} \sin^2 \xi \bigg] 
\end{multline}
\end{widetext}
Here, $\rv_i$ measures the position of $i$th dislocation with respect to the bundle center, and $\xi$ is defined by
\begin{equation}
\cos^{2}\xi=\frac{|\mathbf{\Delta x} _{1 2}|^2}{(R^2-\rho_{1}^{2})(R^2-\rho_{2}^{2})+|\mathbf{\Delta x}_{12}|^2} .
\end{equation}
Due to the presence of the free boundary, this pair potential encodes a significantly more complex dependence on defect orientation and position than the well-known elastic interactions of dislocations in 2D crystals~\cite{halperin_nelson, hirth, nabarro}.  However, we notice the well-known form of logarithmic dislocation interactions in bulk crystals is easily obtained by taking the limit that $\rho_i/R\to 0$ of eq. (\ref{eq: Eint}), for which $\cos \xi \to |\Delta \xv_{12}|/R$ and $\sin \xi \to 1$.  Additionally, we note that when in the limit where either dislocation approaches the boundary, $\rho_i/R \to 1$, the dislocations interactions vanish, which can easily be verified for the case $\cos \xi \to 1$ and $\sin \xi \to 0$.

\subsection{Defect phase diagram of twisted bundles}

Here, we analyze the stability of disclination and dislocations in the cross section of twisted bundles.  As shown previously~\cite{grason_prl_10, grason_pre_11}, and eq. (\ref{eq: disc}), twist-induced stresses couple favorably to the presence of positively charged ($s=+2\pi/6$), 5-fold disclinations, and above a critical threshold of reduced twist, $|(\Omega R)_c= \sqrt{2/9} \simeq 0.47$, this energetic coupling is sufficient to compensate for the positive self-energy cost of a single disclination at any position.  Thus, bundles are unstable to one or more 5-fold disclinations for $|\Omega R| \geq (\Omega R)_c$.  

We consider the stability of a dislocation by considering the energy of a single dislocation, which is polarized by twist-induced stress such that $\bv = b \hat{\phi}$.  Minimizing the sum of eqs. (\ref{eq: Eself}) and (\ref{eq: Etwist}) over radial position, $\rho$, we find the value of twist, $(\Omega R)_*$ at which the net cost of a single dislocation vanishes, $E_{disl}(\rho_*)=0$, where $\rho_*$ is the stable position of the dislocation~\footnote{In the limit of small $R/a$, the stable dislocation necessarily approaches the surface of the bundle where the small $a$ expansion of eq. (\ref{eq: Eself}) fails.   In this limit, in order to resolve $(\Omega R)_*$ it is necessary to maintain the full form of the 5-7 disclination pair energy for a finite $a$.}  For larger bundle twists $|\Omega R| \geq (\Omega R)_*$, one or more dislocations is stable in the low-energy packing of twisted bundles.

\begin{figure}
\center \epsfig{file=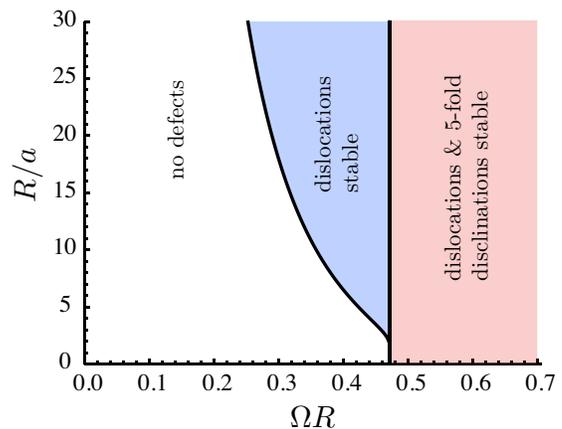, width= 3 in}
\caption{The phase diagram indicating stability of appropriately oriented ($\bv = b \hat{\phi}$) dislocations and 5-fold disclinations in cross sectional order of twisted filament bundles in terms of reduced twist and reduced size of bundles.  }
\label{fig: phase}
\end{figure}

In Fig. \ref{fig: phase} we show the value of both the threshold for disclinations and dislocations, $(\Omega R)_c$ and $(\Omega R)_*$, respectively, as functions of $R/a$, the size of the bundle in units of the lattice spacing, $a \simeq |\bv|$.  The threshold for 5-fold disclinations is independent of bundle size; however, we find that $(\Omega R)_*< (\Omega R)_c$ for all $R/a \geq 2$.  Thus, for fixed bundle size $R/a$, for increasing values of $(\Omega R)$, twisted bundles become unstable to neutral defects, dislocations, before becoming unstable to the 5-fold disclinations in their ground-state packing.

We can roughly estimate the size dependence of $(\Omega R)_*$ in the regime of large bundles.  In this limit, the position of the dislocation is determined by the twist energy alone, which is minimal for $\rho_*= R/\sqrt{3}$.  Solving $E_{disl}(R/\sqrt{3}) =0$ critical twist $(\Omega R)_{*}$, we find 
\begin{equation}
(\Omega R)^{2}_{*}\simeq\frac{\sqrt{3}|b|}{\pi R}\left[\ln\left(\frac{R}{a}\right)-0.072\right] .
\end{equation}
This formula highlights the balance between the logarithmic self-energy of a single dislocation, $\sim K_0 |\bv|^2 \ln (R/a)$, and the compensating dislocation-twist coupling, $\sim - K_0 \Omega^2 R |\bv|$.   Hence, we find that the threshold twist necessary for stabilizing dislocations in the cross section becomes arbitrarily small as bundles become macroscopic in radius, in the $R/a\rightarrow\infty $ limit.  This analysis suggests that dislocations proliferate in large bundles, and therefore, understanding the ground state packing requires the study of multi-dislocation structure and energetics.

\section{Multi-dislocation ground states}

\label{sec: multi}

\subsection{Numerical study}

In this section we explore the structure of multi-dislocation cross sections in the region of intermediate twist, $(\Omega R)_* < |\Omega R| < (\Omega R)_c$.  We base our analysis on a certain class of mechanically stable and high-symmetry dislocation geometries where parallel, $\bv= b \hat{\phi}$ dislocations concentrate along $M$ identical radial lines, or ``arms", spaced evenly at angular intervals of $2 \pi /M$, around the bundle.  A similar class multi-dislocation geometries have been studied in the context of grain-boundary screening of isolated disclinations in 2D crystals~\cite{travesset}.  In these geometries, each radial line of dislocations is line of mirror symmetry in the defect packing so that $\sigma_{r \phi}=0$ along these lines and, by eq. (\ref{eq: PK}), the $\phi$ component of force (the glide direction) vanishes for each dislocation.  The remaining force balance along the radial direction results from repulsive inter-dislocation forces that favor expansion of the array and the Peach-Koehler force on dislocations arising from twist-induced stresses that favors a restoring compression of the array.  As shown in Fig. \ref{fig: lattice}, extended strings of alternating 5- and 7-fold defects constitute tilt grain boundaries across which the orientation of two domains of crystalline order rotates by a discrete angle~\cite{nabarro}.

\begin{figure}
\center
\epsfig{file=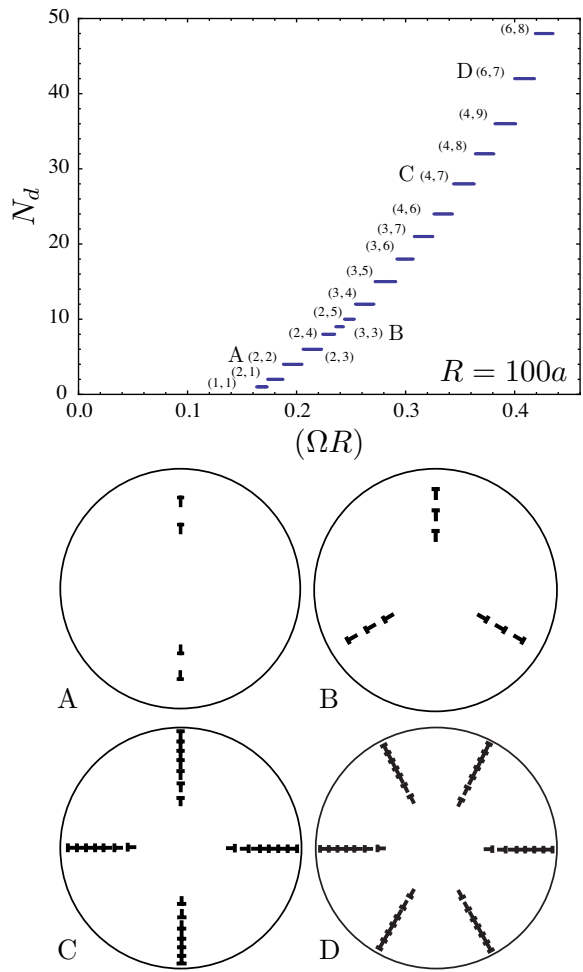, width=3 in}
\caption{ (Top) The total dislocation number for ground-state configurations of twisted bundles with multiple dislocations for a bundle of size, $R=100 a$. Integer pairs, $(M,n)$, refer to the number grain of boundary arms and the number of dislocations per arm, respectively.  (Bottom) A, B, C, and D show 2-,3-, 4- and 6-fold grain boundary geometries, where $\bot$ labels the position of a single dislocation.}
\label{fig: R100}
\end{figure}

To determine the radial position of dislocations in these minimal-energy configurations, we consider the total energy of configurations possessing $N_d$ total disclinations, composed of $M$ equivalent arms of $n= N_d/M$ dislocations per arm.  For a fixed dislocation geometry, reduced twist and bundle size, the sum of the single defect energy, eq. (\ref{eq: Edisl}), and interaction energy between defect pairs, eq. (\ref{eq: Eint}), is numerically minimized with respect to the radial position of the $n$ dislocation ``rings" in the array.  In this analysis, the minimum spacing between successive dislocations along the array is set to be, $a$, the lattice spacing.

\begin{figure}
\center
\epsfig{file=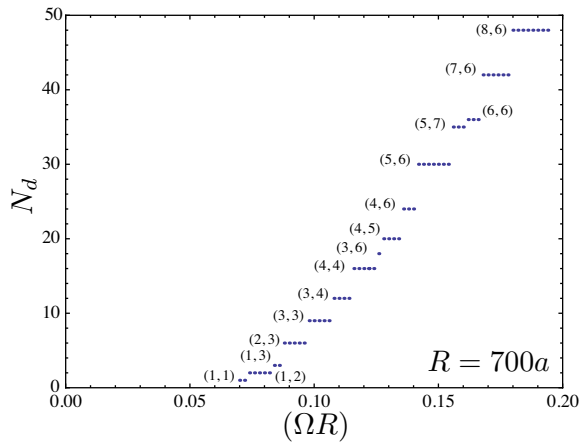, width=3 in}
\caption{ The total dislocation number for ground-state configurations of twisted bundles with multiple dislocations for a bundle of size, $R=700 a$.  Data points are labeled as in Fig. \ref{fig: R100}.}
\label{fig: R700}
\end{figure}

Fig. \ref{fig: R100} shows the results for the number and arrangement of dislocations in a bundle of size $R=100 a$ for a range of twist below the threshold for stable 5-fold defects.  As the twist increases beyond $(\Omega R)_* \simeq 0.16$, the number of dislocations favored in the cross section increases quickly.  For each value of $N_d$, the geometry of the dislocation packing is labeled by the integer pair, $(M,n)$, denoting the number of grain boundary arms and the number of dislocations per arm, respectively.   Along with the total dislocation number, the number of radial grain boundaries also grows with $(\Omega R)$, leading 2-, 3-, 4- and 6-fold grain boundary geometries depicted in Fig. \ref{fig: R100}.  

In Fig. \ref{fig: R700} we show results for $N_d$ vs. $(\Omega R)$ for a much larger bundle, $R= 700 a$.  While this bundle shows a similar trend with increasing twist, we note that the threshold for stable dislocations is markedly reduced, $(\Omega R)_* \simeq 0.07$ and a distinct sequence of dislocation geometries is predicted as $N_d$ increases rapidly with twist.   Notably, we find for all multi-dislocation geometries over a range from $R/a = 20$ to $700$,  that grain-boundary arms penetrate only a fraction of the distance from the bundle surface to the bundle center, terminating in the bulk at a finite radius, a feature uncommon in bulk crystalline materials.

\begin{figure}
\center
\epsfig{file=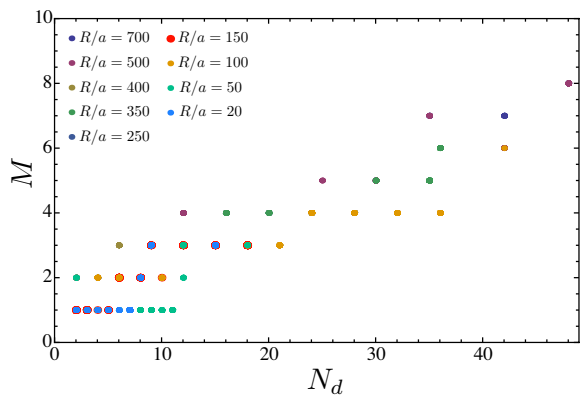, width=3 in}
\caption{Plot of number of grain boundarm arms, $M$, vs. total dislocation number, $N_d$ for bundle sizes in the range $R/a=20-700$.}
\label{fig: M}
\end{figure}

To investigate the evolution of grain-boundary structure in bundles with increasing twist, in Fig. \ref{fig: M} we plot the number of grain boundary arms, $M$, vs. $N_d$ for all values of $R/a$ studied. Over the range of dislocations explored here (up to $N_d = 50$) we find little systematic dependence of the growth in the number of grain boundaries on bundle size.  Despite more than an order of magnitude variation of bundle size, the trend of increasing number of grain boundaries is consistent with a roughly linear relationship, $M\sim N_d$ for all $R/a$.  This suggests that the optimal dislocation geometry is nominally determined by $N_d$ alone, which in turn is regulated by $\Omega R$ in a manifestly size-dependent manner, as evidenced by the results for $R=100 a$ and $R= 700a$ bundles, of Figs. \ref{fig: R100} and \ref{fig: R700}, respectively.

\subsection{Scaling analysis}
Above, we found that the gross structure of the multi-dislocation ground state is predominantly sensitive to the total dislocation number.  Here, we consider a simple scaling argument to understand the dependence of $N_d$ on bundle twist and size.  This argument is similar to the geometric analysis of ``grain-boundary scars" on spherical crystals~\cite{bowick_travesset_nelson}, with the notable exception that in the present case, neutral grain boundaries form in the absence of excess point disclinations.  According to the compatibility relation, eq. (\ref{eq: compat}), we can formally consider the source of  twist-induced stresses, $K_{T}=3 \Omega^2$, to be a uniform areal density of negatively charged disclinations.   Integrating this charge density over the cross section of the bundle we define an effective disclination charge,
\begin{equation}
s_{eff} = - 3 \pi (\Omega R)^2.
\end{equation}
As argued in ref. \cite{bowick_travesset_nelson}, the strain generated by this ``topological defect" can be compensated by the presence of $M$ radial grain boundaries, each of which accommodates a rotation of $\theta \simeq a/d$, where $d$ is the mean dislocation spacing along the boundary~\cite{halperin_bruinsma}.  Equating the effective topological charge to the total grain boundary rotation, we find the mean-spacing between dislocations,
\begin{equation}
d^{-1} \approx a^{-1} (\Omega R)^2 /M.
\end{equation}
Integrating the linear density of dislocations along the length of grain boundaries ($\sim R$) we find the mean number of dislocations per arm,
\begin{equation}
n \approx (R/a) (\Omega R)^2/M .
\end{equation}
Multiplying $n$ by the number of grain boundaries in the cross section, we argue that for large twist, $N_d \sim (R/a) (\Omega R)^2$.    To capture both limiting cases of large twist and the critical twist at which $N_d$ vanishes, we construct the following scaling form for total disclination number,
\begin{equation}
\label{eq: scaling}
N_d \sim (R/a) \big[ (\Omega R)^2 - (\Omega R)^2_* \big] .
\end{equation}
Hence, not only do larger bundles become unstable to dislocations at smaller values of bundle twist, the growth of the optimal number of dislocations with ``excess" twist in large bundles is also more rapid than in smaller bundles.

\begin{figure}
\center
\epsfig{file=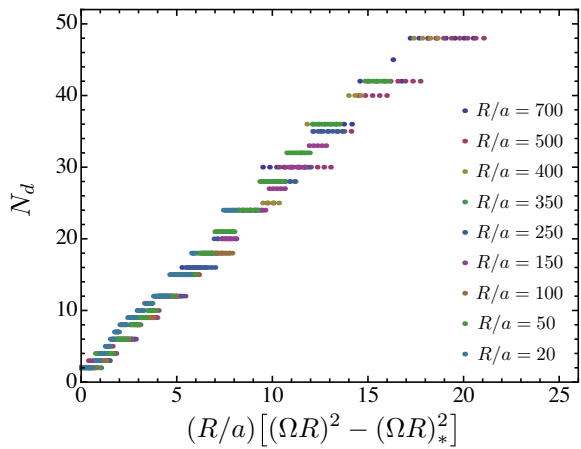, width=3 in}
\caption{Plot of dislocation number vs. scaling prediction, eq. (\ref{eq: scaling}), for bundle sizes in the range $R/a=20-700$.}
\label{fig: scaling}
\end{figure}

In Fig. \ref{fig: scaling} we compare the total dislocation number of the numerically-determined ground states to the scaling prediction, eq. (\ref{eq: scaling}).   We find that the scaling prediction agrees well with numerical results over the range of $N_d$ and the large range of bundle sizes explored here, $20 \leq R/a \leq 700$.  

\section{Discussion}

In summary, we have shown that geometric frustration arising from helical twist in two-dimensionally ordered filament bundles restructures the ground-state packing at intermediate twist by favoring the presence of appropriately oriented, edge dislocations.  Based on the continuum theory of disclinations and dislocations in filament bundles, we show that {\it dislocations} become favorable in the cross section at twist smaller than the critical twist needed to stabilize {\it 5-fold disclinations}.  Unlike the case of stable 5-fold disclinations studied previously~\cite{grason_prl_10, grason_pre_11}, here we find that the threshold for stable dislocations in bundles is highly dependent on the size of the bundle compared to the microscopic inter-filament spacing.  Above the threshold single dislocations in twisted bundles, we predict a rich spectrum of low-symmetry ground-state order.  Twist-induced stresses in filament bundles lead to a natural tendency to ``polygonalize" the cross-sectional packing, giving rise to low-energy structures where multiple crystalline domains are separated by radially-extending grain boundaries that terminate in the bundle bulk, not unlike the finite-length grain-boundary scars of spherical crystals.

This study is significant in the context of frustrated order because it demonstrates that ``neutral" configurations of topological defects may effectively screen ``charged", geometrically-induced stresses, like the stresses generated by filament twist in bundles.   Previous studies of defects on curved, crystalline membranes, have predicted extended chains dislocations, or scars, only in the presence of excess disclinations that are themselves either forced in by topology~\cite{bowick_travesset_nelson, bowick_caccuito} or, in the case of a membranes with a free boundary, as the result of energetic coupling to curvature-induced stresses~\cite{giomi_bowick, yao}.  In these cases, the dislocation arrays function to screen the disclination stress more efficiently than the stresses induced by Gaussian curvature.  Here, we show that neutral arrays of 5-7 disclination pairs flood the ground-state packing of crystalline bundles well before twist favors the incorporation of excess 5-fold defects.  That is, dislocations arrays are also driven into the packing of frustrated materials by the tendency to screen geometrically-induced stresses alone.  Due to the formal relationship between the non-linear elasticity of twisted bundles and curved, crystalline membranes, we expect that the novel grain-boundary geometries predicted here may also occur as ground states of the latter system.  

In the context of filamentous materials, the present study is significant for two reasons.  First, it identifies $\Omega R$ and $R/a$ as the two geometric parameters that govern the ground state packing of helically-twisted bundles.  Importantly, we show that the critical degree of twist at which the bundle cross section becomes unstable to topological defects is crucially sensitive to bundle size, $R/a$.  We may relate the reduced twist of the bundle to the tilt angle, $\theta_{max}$ of the outermost filament with respect the pitch axis of the helical bundle by, $\theta_{max} = \tan^{-1} (\Omega R)$.  For small bundles, less than a few radial filament layers, the critical twist for stability of {\it any} defect type in the cross section corresponds to a degree of tilt greater than $25^\circ$.  Because the critical twist decreases with bundle size as $(\Omega R)_* \sim a/R \ln (R/a)$, for bundles that are macroscopically large compared to filament size, say $R=100 a$, the critical degree of filament tilt is markedly reduced to nearly $9^\circ$.  For comparison, we note the helical twist of certain collagen fibrils is in the range $15-17^\circ$ of helical tilt~\cite{wess, ottani}.

The case of collagen points to the second important result regarding the structure of optimally-packed twisted fibers.  In many tissue types, collagen fibrils form with lateral dimensions hundreds of times larger than the roughly 1 nm scale of constituent filaments, making multiple dislocations energetically favorable even in fibrils of relatively modest twist.  Notably the precise nature of the cross-sectional ordering of collagen molecules in fibrils is a long-standing and open question, in part, due to small-angle scattering data that suggest cross sections are composed of unknown superpositions of crystalline inter-molecular order and disordered inter-molecular packing of some type.  Numerous models have been proposed to infer the real-space packing~\cite{prockop}, many of which mix aspects of crystalline and non-crystalline order in novel ways~\cite{sadoc_charvolin}.  To date, the model most consistent with observed features of x-ray scattering data was proposed by Hulmes, Wess, Prockop and Fratzl~\cite{hulmes}.  In this model, multiple crystalline domains in a cylindrical bundles are separated by grain boundaries extending radially from a central, low-density region to the surface of the fibril.  Remarkably, this model is very similar in gross structure to the ground states of twisted bundles predicted for large bundles of intermediate twist (for example in Fig. \ref{fig: R100}).  Though this model of collagen fibril packing did not take into account the effects of twist explicitly, we believe many of the key features of the ``disordered" packing of these materials may be understood as crucial elements of energy-minimizing packings of twisted bundles.  Future work will explore the form factor of ideal packings of twisted bundles and critically test the intriguing and putative connection between defects in the ground states of twisted bundles and the disorder in the collagen fibril.

\begin{acknowledgments}
The authors would like to thank I. Bruss for a careful reading of this manuscript.  This work was supported by the NSF Career program under DMR Grant 09-55760.  
\end{acknowledgments}

\end{document}